\documentclass[12pt]{iopart}
\usepackage{iopams}
\usepackage{setstack}
\usepackage{psfig}

\newcommand{\half}{\mbox{$\textstyle \frac{1}{2}$}}
\newcommand{\cP}{\mathcal P}
\newcommand{\cC}{\mathcal C}
\newcommand{\cT}{\mathcal T}
\begin{document}

\title[Calculation of the Hidden Symmetry Operator for a $\cP\cT$-Symmetric
Square Well]{Calculation of the Hidden Symmetry Operator for a
$\cP\cT$-Symmetric Square Well}

\author[Bender and Tan]{Carl~M~Bender$^{*}$ and Barnabas~Tan$^{\dagger}$}

\address{${}^{*}$Department of Physics, Washington University, St. Louis MO
63130, USA}

\address{${}^{\dagger}$Blackett Laboratory, Imperial College, London SW7 2BZ,
UK}

\begin{abstract}
It has been shown that a Hamiltonian with an unbroken $\cP\cT$ symmetry also
possesses a hidden symmetry that is represented by the linear operator $\cC$.
This symmetry operator $\cC$ guarantees that the Hamiltonian acts on a Hilbert
space with an inner product that is both positive definite and conserved in
time, thereby ensuring that the Hamiltonian can be used to define a unitary
theory of quantum mechanics. In this paper it is shown how to construct the
operator $\cC$ for the $\cP\cT$-symmetric square well using perturbative
techniques.
\end{abstract}

\submitto{\JPA}


\section{Introduction}
\label{s1}

The discovery \cite{r1,r2,r3} that there were huge classes of $\cP\cT$-symmetric
non-Hermitian Hamiltonians of the form $H=p^2+x^2(ix)^\epsilon$ ($\epsilon\geq
0$) whose spectra were real and positive led to the investigation of many new
kinds of $\cP\cT$-symmetric model Hamiltonians. One particularly elegant model
is the $\cP\cT$-symmetric square well, whose Hamiltonian on the domain $0<x<\pi$
is given by
\begin{equation}
H=p^2+V(x),
\label{e1}
\end{equation}
where $V(x)=\infty$ for $x<0$ and $x>\pi$ and 
\begin{equation}
V(x) = \left\{
\begin{array}{cl}
i\epsilon \quad & \mbox{for $\frac{\pi}{2}<x<\pi,$} \\
-i\epsilon \quad & \mbox{for $0<x<\frac{\pi}{2}.$} \\
\end{array}
\right.
\label{e2}
\end{equation}
This Hamiltonian reduces to the conventional Hermitian square well in the limit
as $\epsilon\to0$. For $H$ in (\ref{e1}) the parity operator $\cP$ performs a
reflection about $x=\frac{\pi}{2}$: $\cP:x\to\pi-x$. The $\cP\cT$-symmetric
square-well Hamiltonian was invented and first examined by Znojil \cite{r4} and
it has been heavily studied by many other researchers \cite{r5,r6,r7,r8}.

The principal challenge in understanding non-Hermitian $\cP\cT$-symmetric
Hamiltonians was to show that they describe unitary time evolution. This was
accomplished by the discovery of a hidden symmetry operator called $\cC$. This
operator is used to define the Hilbert-space inner product with respect to which
the Hamiltonian is self-adjoint \cite{r9}. In Ref.~\cite{r9} the $\cC$ operator
in coordinate space was shown to have a representation as a sum over the
eigenfunctions $\phi_n(x)$ of the Hamiltonian:
\begin{equation}
\cC(x,y)=\sum_{n=0}^\infty\phi_n(x)\phi_n(y),
\label{e3}
\end{equation}
where the eigenfunctions are normalized so that they are eigenstates of the
$\cP\cT$ operator with eigenvalue 1,
\begin{equation}
\cP\cT\phi_n(x)=\phi_n(x),
\label{e4}
\end{equation}
and the integral of the square of the $n$th eigenfunction oscillates in sign:
\begin{equation}
\int dx\,[\phi_n(x)]^2=(-1)^n.
\label{e5}
\end{equation}

The discovery of the $\cC$ operator led immediately to attempts to calculate it
for various model Hamiltonians. For the elementary $\cP\cT$-symmetric
non-Hermitian Hamiltonian $H=\half p^2+\half x^2+ix$, the exact $\cC$ operator
is given by \cite{r10}
\begin{equation}
\cC=e^{-2p}\cP.
\label{e6}
\end{equation}
However, for more complicated Hamiltonians the $\cC$ operator cannot be obtained
in closed form. It was shown in Ref.~\cite{r11} how to use perturbative methods
to evaluate the sum in (\ref{e3}) for the Hamiltonian
\begin{equation}
H=p^2+x^2+i\epsilon x^3.
\label{e7}
\end{equation}
In Ref.~\cite{r12} this perturbative procedure was extended to
quantum-mechanical Hamiltonians having several degrees of freedom.

The perturbative methods used in Refs.~\cite{r11} and \cite{r12} were not
powerful enough to be used in quantum field theory, so a simple recipe for
finding $\cC$ was devised that can be used in systems having an infinite number
of degrees of freedom \cite{r10}. The procedure was to solve the three
simultaneous algebraic equations satisfied by $\cC$:
\begin{equation}
\cC^2=1,\quad [\cC,\cP\cT]=0,\quad [\cC,H]=0.
\label{e8}
\end{equation}
This recipe gives the $\cC$ operator as a product of the exponential of an
antisymmetric Hermitian operator $Q$ and the parity operator $\cP$:
\begin{equation}
\cC=e^Q\cP.
\label{e9}
\end{equation}
Note that $Q=-2p$ for the $\cC$ operator in (\ref{e6}). Mostafazadeh has shown
that the square root of the positive operator $e^Q$ can be used to construct a  
similarity transformation that converts a non-Hermitian $\cP\cT$-symmetric
Hamiltonian $H$ to an equivalent Hermitian Hamiltonian $h$ \cite{r13}: $h=e^{-Q/
2}He^{Q/2}$.

In all the examples studied so far the $\cC$ operator is a combination of
integer powers of $x$ and integer numbers of derivatives multiplying the parity
operator $\cP$. Hence, the $Q$ operator is a polynomial in the operators $x$ and
$p=-i\frac{d}{dx}$. The novelty of the $\cP\cT$-symmetric square-well
Hamiltonian (\ref{e1}-\ref{e2}) is that $\cC$ contains {\it integrals} of $\cP$
and thus the $Q$ operator, while it is a simple function, is {\it not} a
polynomial in $x$ and $p$ and therefore cannot be found easily by the algebraic
perturbative methods that were introduced in Ref.~\cite{r10}. Thus, in
Sec.~\ref{s2} we calculate $\cC$ for this Hamiltonian by using the perturbative
techniques that were devised in Ref.~\cite{r11}. In Sec.~\ref{s3} we make some
concluding remarks.

\section{Perturbative calculation of the $\cC$ operator}
\label{s2}

The procedure we use here is as follows: First, we solve the Schr\"odinger
equation
\begin{equation} 
-\phi_n''(x)+V(x)\phi_n(x)=E_n\phi_n(x)\quad(n=0,\,1,\,2,\,3,\,\dots)
\label{e10}
\end{equation}
subject to the boundary conditions $\phi_n(0)=\phi_n(\pi)=0$. We obtain the
eigenfunction $\phi_n(x)$ as a perturbation series to second order in powers of
$\epsilon$. The eigenfunctions are then normalized according to (\ref{e4}) and
(\ref{e5}). Next, we substitute the eigenfunctions into the formula (\ref{e3})
and evaluate the sum. The advantage of the domain of the square well being $0<x<
\pi$ is that this sum reduces to a set of Fourier sine and cosine series that
can be evaluated in closed form. After evaluating the sum, it is convenient to
translate the domain of the square well to the more symmetric region $-\frac{\pi
}{2}<x<\frac{\pi}{2}$. On this domain the parity operator in coordinate space is
$\cP(x,y)=\delta(x+y)$. Finally, we show that the $\cC$ operator to order
$\epsilon^2$ has the form in (\ref{e9}), and we evaluate the function $Q$ to
order $\epsilon^2$. Our final result for $Q(x,y)$ on the domain $-\frac{\pi}{2}<
x<\frac{\pi}{2}$ is
\begin{equation}
Q(x,y)=\textstyle{\frac{1}{4}}i\epsilon[x-y+\varepsilon(x-y)\,(|\,x+y\,|-\pi)]+
\mathcal{O}(\epsilon^3),
\label{e11}
\end{equation}
where $\varepsilon(x)$ is the standard step function
\begin{equation}
\varepsilon(x)=\left\{
\begin{array}{cl}
1\quad&\mbox{($x>0$),}\\
0\quad&\mbox{($x=0$),}\\
-1\quad&\mbox{($x<0$).}\\
\end{array}
\right.
\label{e12}
\end{equation}

\subsection{Solution of the Schr\"odinger Equation}
We begin our analysis by solving the Schr\"odinger equation (\ref{e10}) in the
right ($x>\frac{\pi}{2}$) and left ($x<\frac{\pi}{2}$) regions of the square
well:
\begin{eqnarray}
&&\phi_{n,\,{\rm R}}(x)=a_n\,\bigg\{i^{\frac{1}{2}(1-(-1)^{n})}\sin (n+1)x\nonumber\\
&&\hspace{-2.4cm}+\left[i^{\frac{1}{2}(1+(-1)^{n})}\left(\frac{\pi}{2}-\frac{x}{2}
\right)\frac{(-1)^n\cos(n+1)x}{(n+1)}-\frac{1}{2}(1-(-1)^{n})\frac{\sin(n+1)x}
{2(n+1)^{2}}\right]\epsilon\nonumber\\
&&\hspace{-2.4cm}+\,i^{\half(1-(-1)^{n})}\left[\half(1+(-1)^{n})\left(\frac{x}{4
}-\frac{\pi}{4}\right)\frac{\cos(n+1)x}{(n+1)^3}+\left(\frac{x^2}{8}-\frac{\pi x
}{4}+\frac{\pi^2}{16}\right)\frac{\sin(n+1)x}{(n+1)^{2}}\right]\epsilon^2
\nonumber\\
&&\hspace{-2.4cm}+\mathcal{O}(\epsilon^3)\bigg\}\qquad\mbox{($x>\frac{\pi}{2}$
),}
\label{e13}\\
\nonumber\\
&&\phi_{n,\,{\rm L}}(x)=a_n\,\bigg\{i^{\half(1-(-1)^{n})}\sin(n+1)x\nonumber\\
&&\hspace{-2.4cm}+\left[i^{\half(1+(-1)^{n})}\frac{x}{2}\frac{(-1)^n\cos(n+1)x}
{(n+1)}+\,\half(1-(-1)^{n})\frac{\sin(n+1)x}{2(n+1)^{2}}\right]\epsilon
\nonumber\\
&&\hspace{-2.4cm}+\,i^{\half(1-(-1)^{n})}\left[\frac{1}{2}(1+(-1)^{n})\frac{x}{4
}\frac{\cos(n+1)x}{(n+1)^3}+\left(\frac{x^2}{8}-\frac{\pi^2}{16}\right)\frac{
\sin(n+1)x}{(n+1)^2}\right]\epsilon^2\nonumber\\
&&\hspace{-2.4cm}+\mathcal{O}(\epsilon^3)\bigg\}\qquad\mbox{($x<\frac{\pi}{2}$
).}
\label{e14}
\end{eqnarray}
These eigenfunctions and their first derivatives are continuous at $x=\frac{\pi}
{2}$. 

Having found the eigenfunction $\phi_n(x)$ to second order in $\epsilon$, we
can give the formula for the corresponding eigenvalues:
$$E_n=(n+1)^2+\frac{(-1)^n[2-(-1)^n]}{4(n+1)^2}\epsilon^2+\mathcal{O}
(\epsilon^4).$$
However, these eigenvalues are not needed for calculating the $\cC$ operator.

\subsection{Normalization of the Eigenfunctions}
The normalization requirements in (\ref{e4}-\ref{e5}) give the value of the
coefficient $a_n$ in (\ref{e13}-\ref{e14}):
\begin{equation}
\hspace{-1.2cm}a_n=\sqrt{\frac{2}{\pi}}\left[1-(-1)^n\left(\frac{(2-(-1)^n)}
{(6-2(-1)^n)(n+1)^4}-\frac{(-1)^n\pi^2}{16(n+1)^2}\right)\epsilon^2
+\mathcal{O}(\epsilon^4)\right].
\label{e15}
\end{equation}
With this normalization, the $\cP\cT$ inner product between $\phi_m(x)$ and
$\phi_n(x)$ is $(-1)^n\delta_{mn}+\mathcal{O}(\epsilon^4)$.

\subsection{Calculation of $\cC(x,y)$ to Leading Order (Zeroth Order) in
$\epsilon$}
The next step is to construct the operator $\cC(x,y)$, which is given in
(\ref{e3}) as a sum, by directly substituting the eigenfunctions $\phi_n(x)$
from (\ref{e13}-\ref{e14}). In general, there are four different regions of $x$
and $y$ to consider:
\begin{enumerate}
\item $x>\frac{\pi}{2}$, $y>\frac{\pi}{2},$
\item $x>\frac{\pi}{2}$, $y<\frac{\pi}{2},$
\item $x<\frac{\pi}{2}$, $y<\frac{\pi}{2},$
\item $x<\frac{\pi}{2}$, $y>\frac{\pi}{2}.$
\end{enumerate}
However, to zeroth-order in $\epsilon$, $\phi_n$ is common to all four regions
and the calculation is easy. We find that
\begin{equation}
\cC^{(0)}(x,y)=\frac{2}{\pi}\sum_{n=0}^{\infty}\,(-1)^n\sin(n+1)x\,\,\sin(n+1)y.
\label{e16}
\end{equation}
This is just the Fourier sine series for the parity operator in the range
$0<x<\pi$:
\begin{equation}
\cC^{(0)}(x,y)=\delta(x+y-\pi).
\label{e17}
\end{equation}
On the symmetric domain $-\frac{\pi}{2}<x<\frac{\pi}{2}$ this formula becomes
\begin{equation}
\cC^{(0)}(x,y)=\delta(x+y),
\label{e18}
\end{equation}
which is equivalent to the coordinate-space condition of completeness.

\subsection{Calculation of $\cC(x,y)$ to First Order in $\epsilon$}
The calculation of $\cC(x,y)$ to first order in $\epsilon$ requires the
evaluation of Fourier sine and cosine series. These are expressed in terms of
single and double integrals of delta functions. Here, we describe the
calculation for the region $x>\frac{\pi}{2}$, $y>\frac{\pi}{2}$. The calculation
for the other three regions is similar.

From (\ref{e13}) and (\ref{e15}) the first-order contribution to $\cC(x,y)$ is
\begin{eqnarray}
\hspace{-0.3cm}\cC^{(1)}(x,y)&=&\frac{1}{\pi}\sum_{n=0}^\infty\left[(\pi-x)\frac
{i(-1)^n}{n+1}\cos[(n+1)x]\,\sin[(n+1)y]\right]\nonumber\\
&&+\frac{1}{\pi}\sum_{n=0}^\infty\left[(\pi-y)\frac{i(-1)^n}{n+1}\sin[(n+1)
x]\,\cos[(n+1)y]\right]\nonumber\\
&&-\frac{1-(-1)^n}{\pi}\sum_{n=0}^\infty\left[i^{\frac{1}{2}(1-(-1)^n)}
\frac{\sin[(n+1)x]\,\sin[(n+1)y]}{(n+1)^2}\right].
\label{e19}
\end{eqnarray}
The first two terms of $\cC^{(1)}$ can be expressed as single integrals of the
parity operator, with the first having the upper limit $x$,
\begin{eqnarray}
\int_{\frac{\pi}{2}}^x dt\,\delta(t+y-\pi)&=&-\frac{2}{\pi}\sum_{n=0}^\infty
\left[\frac{(-1)^n}{n+1}\cos[(n+1)x]\,\sin[(n+1)y]\right]\nonumber\\
&&+\frac{1}{\pi}\sum_{n=0}^\infty\left[\frac{(-1)^n}{n+1}\sin[(2n+2)y]\right],
\label{e20}
\end{eqnarray}
and the second having the upper limit $y$,
\begin{eqnarray}
\int_{\frac{\pi}{2}}^y dt\,\delta(x+t-\pi)&=&-\frac{2}{\pi}\sum_{n=0}^\infty
\left[\frac{(-1)^n}{n+1}\sin[(n+1)x]\,\cos[(n+1)y]\right]\nonumber\\
&&+\frac{1}{\pi}\sum_{n=0}^\infty\left[\frac{(-1)^n}{n+1}\sin[(2n+2)x]\right].
\label{e21}
\end{eqnarray}

The third term of $\cC^{(1)}$ involves sines of even $x$ and $y$. To obtain a
series of sines of even $x$ and $y$, we subtract the series represented by the
parity operator $\delta(x+y-\pi)$ from the series represented by $\delta(x-y)$.
The factor $(n+1)^2$ in the denominator implies that the third term of $\cC^{(1)
}$ may be expressed as a double integral. To maintain the symmetry, we note that
a double integral with respect to $x$ contributes half of the third term of
$\cC^{(1)}$, while a double integral with respect to $y$ contributes the other
half.

This analysis allows us to express $\cC^{(1)}$ in the region $x>\frac{\pi}{2}$,
$y>\frac{\pi}{2}$ in terms of single integrals of the parity operator and double
integrals in $x$ and $y$:
\begin{eqnarray}
\cC^{(1)}(x,y)&=&i\,\Bigg\{\left(\frac{x}{2}-\frac{\pi}{2}\right)\int_{\pi/2}^x
dt\,\delta(t+y-\pi)\nonumber\\
&&+\frac{1}{4}\int_{\pi/2}^x dt\int_{\pi/2}^t ds\,\Big[\delta(s-y)-\delta(s+y-
\pi)\Big]\nonumber\\
&&+\frac{x}{4}-\frac{\pi}{4}+(x\leftrightarrow y)\Bigg\}.
\label{e22}
\end{eqnarray}
We simplify this result by evaluating integrals over delta functions and obtain
\begin{equation}
\cC^{(1)}(x,y)={\textstyle\frac{1}{4}}i(|x-y|+x+y-2\pi)\quad
\left(x>\textstyle{\frac{\pi}{2}},\,y>\textstyle{\frac{\pi}{2}}\right).
\label{e23}
\end{equation}

\begin{figure}[th]\vspace{3.9in}
\includegraphics{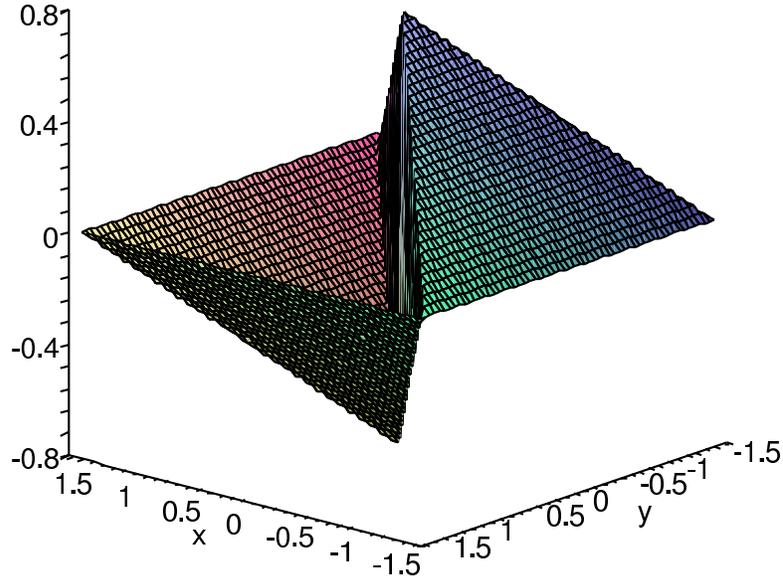}
\caption{Three-dimensional plot of the imaginary part of $\cC^{(1)}(x,y)$, the
first-order perturbative contribution in (\ref{e27}) to the $\cC$ operator in
coordinate space. The plot is on the symmetric square domain $-\frac{\pi}{2}<(x,
y)<\frac{\pi}{2}$. Note that $\cC^{(1)}(x,y)$ vanishes on the boundary of this
square domain because the eigenfunctions $\phi_n(x)$ in (\ref{e3}) are required
to vanish at $x=0$ and $x=\pi$.}
\label{f1}
\end{figure}

The calculation of $\cC^{(1)}(x,y)$ for the remaining three regions follows a
similar procedure and we get
\begin{eqnarray}
\cC^{(1)}(x,y)&=&\nonumber\\
&&\hspace{-1.9cm}\left\{
\begin{array}{ll}
\half i[(x-\pi)\,\theta(x+y-\pi)+y\,\theta(\pi-x-y)]\quad&\mbox{$\left(x>\frac{
\pi}{2},\,y<\frac{\pi}{2}\right),$}\\
{\textstyle\frac{1}{4}}i[-|x-y|+(x+y)]\quad&\mbox{$\left(x<\frac{\pi}{2},\,y<
\frac{\pi}{2}\right),$}\\
\half i[(y-\pi)\,\theta(x+y-\pi)+x\,\theta(\pi-x-y)]\quad&\mbox{$
\left(x<\frac{\pi}{2},\,y>\frac{\pi}{2}\right),$}\\
\end{array}\right.
\label{e24}
\end{eqnarray}
where $\theta(x)$ is the Heaviside step function,
\begin{equation}
\theta(x)=\left\{
\begin{array}{cl}
1\quad &\mbox{($x\geq0$),} \\ 0\quad & \mbox{($x<0$).}\\
\end{array}\right.
\label{e25}
\end{equation}
Finally, we condense the four expressions for $\cC^{(1)}(x,y)$ in the four
different regions into a single expression:
\begin{eqnarray}
\cC^{(1)}(x,y)&=&{\textstyle\frac{1}{4}}i[x+y-\pi-\theta(\pi-x-y)\,(|x-y|-\pi)
\nonumber\\
&&+\theta(x+y-\pi)\,(|x-y|-\pi)].
\label{e26}
\end{eqnarray}
On the symmetric region $-\frac{\pi}{2}<(x,y)<\frac{\pi}{2}$, this expression
becomes
\begin{eqnarray}
\cC^{(1)}(x,y)&=&{\textstyle\frac{1}{4}}i[x+y+\varepsilon(x+y)\,(|x-y|-\pi)].
\label{e27}
\end{eqnarray}
We plot the imaginary part of $\cC^{(1)}$ in Fig.~\ref{f1} as a function of
$x$ and $y$.

\subsection{Calculation of $\cC(x,y)$ to Second Order in $\epsilon$}

The procedure for calculating $\cC^{(2)}(x,y)$ is similar to that used for
calculating $\cC^{(1)}(x,y)$, albeit more tedious. We must calculate sums of
products of sines and cosines, but this time the presence of factors of $(n+1
)^4$, $(n+1)^3$ and $(n+1)^2$ in the denominator requires the use of quadruple,
triple, and double integrals of delta functions to simplify the expression for
$\cC^{(2)}(x,y)$. We discuss the calculation of $\cC^{(2)}(x,y)$ explicitly for
the region $x<\frac{\pi}{2}$, $y<\frac{\pi}{2}$. The calculation of $\cC^{(2)}
(x,y)$ for the other three regions is similar.

From (\ref{e13}) and (\ref{e15}) we see that the second-order calculation of
$\cC(x, y)$ gives
\begin{eqnarray}
\cC^{(2)}(x,y)&=&\frac{2}{\pi}\sum_{n=0}^\infty\left[\frac{x}{4}\,\frac{\cos
[(2n+1)x]\,\sin[(2n+1)y]}{(2n+1)^3}\right]\nonumber\\
&&+\frac{2}{\pi}\sum_{n=0}^\infty\left[\frac{y}{4}\,\frac{\sin[(2n+1)x]\,\cos[
(2n+1)y]}{(2n+1)^3}\right]\nonumber\\
&&+\frac{2}{\pi}\sum_{n=0}^\infty\left[\left(\frac{x^2}{8}+\frac{y^2}{8}\right)
\frac{(-1)^n\sin[(n+1)x]\,\sin[(n+1)y]}{(n+1)^2}\right]\nonumber\\
&&-\frac{2}{\pi}\sum_{n=0}^\infty\left[\frac{x}{4}\,\frac{\cos[(2n+2)x]
\,\sin[(2n+2)y]}{(2n+2)^3}\right]\nonumber\\
&&-\frac{2}{\pi}\sum_{n=0}^\infty\left[\frac{y}{4}\,\frac{\sin[(2n+2)x]
\,\cos[(2n+2)y]}{(2n+2)^3}\right]\nonumber\\
&&-\frac{2}{\pi}\sum_{n=0}^\infty\left[\frac{1}{2}\,\frac{\sin[(2n+1)x]
\,\sin[(2n+1)y]}{(2n+1)^4}\right]\nonumber\\
&&-\frac{2}{\pi}\sum_{n=0}^\infty\left[\frac{1}{2}\,\frac{\sin[(2n+2)x]
\,\sin[(2n+2)y]}{(2n+2)^4}\right]\nonumber\\
&&-\frac{2}{\pi}\sum_{n=0}^\infty\left[\frac{xy}{4}\,\frac{(-1)^n\cos[(n+1)x]\,
\cos[(n+1)y]}{(n+1)^2}\right].
\label{e28}
\end{eqnarray}
We have been able to evaluate each of these Fourier series exactly and to
express the result as multiple integrals over delta functions:
\begin{eqnarray}
\cC^{(2)}(x,y)&=&-{\textstyle\frac{1}{8}}x^2\int_x^{\pi/2}dt\int_t^{\pi/2}ds\,
\delta(s+y-\pi)\nonumber\\
&&-{\textstyle\frac{1}{4}}x\int_x^{\pi/2}dt\int_t^{\pi/2}ds\int_s^{\pi/2}dr\,
\delta(r+y-\pi)\nonumber\\
&&-{\textstyle\frac{1}{4}}\int_x^{\pi/2}dt\int_t^{\pi/2}ds\int_s^{\pi/2}dr
\int_r^{\pi/2}dp\,\delta(p-y)\nonumber\\
&&-{\textstyle\frac{1}{8}}xy\int_y^{\pi/2}dt\int_x^{\pi/2}ds\,\delta(s+t-\pi)+
{\textstyle\frac{1}{8}}x^2y\nonumber\\
&&-{\textstyle\frac{1}{16}}xy\pi+{\textstyle\frac{1}{24}}x^3+(x\leftrightarrow
y).
\label{e29}
\end{eqnarray}
Evaluating these integrals gives $\cC^{(2)}(x,y)$ for the region $x<\frac{\pi}
{2}$, $y<\frac{\pi}{2}$:
\begin{equation}
\cC^{(2)}(x,y)=-\textstyle{\frac{1}{24}}|x-y|^3+\textstyle{\frac{1}{8}}x^2y+
\textstyle{\frac{1}{24}}x^3-\textstyle{\frac{1}{16}}x\pi y+(x\leftrightarrow y).
\label{e30}
\end{equation}

\begin{figure}[th]\vspace{3.9in}
\includegraphics{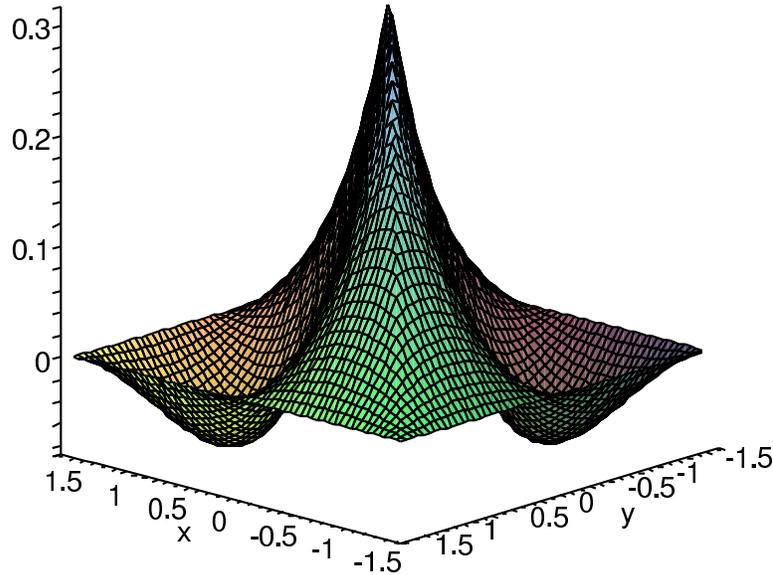}
\caption{Three-dimensional plot of $\cC^{(2)}(x,y)$ in (\ref{e31}) on the
symmetric square domain $-\frac{\pi}{2}<(x,y)<\frac{\pi}{2}$. The function
$\cC^{(2)}(x,y)$ vanishes on the boundary of this square domain because the
eigenfunctions $\phi_n(x)$ from which it was constructed vanish at the
boundaries of the square well.}
\label{f2}
\end{figure}

The calculation of $\cC^{(2)}(x, y)$ for the remaining three regions follows a
similar procedure. Combining the contributions from the four regions and
transforming to the symmetric domain $-\frac{\pi}{2}<(x,y)<\frac{\pi}{2}$,
we obtain the single expression:
\begin{eqnarray}
\hspace{-0.6cm}\cC^{(2)}(x,y)&=&{\textstyle\frac{1}{96}}\pi^3+{\textstyle\frac{
1}{8}}xy\pi-{\textstyle\frac{1}{16}}\pi^2(x+y)\,\varepsilon(x+y)+{\textstyle
\frac{1}{8}}\pi(x|x|+y|y|)\,\varepsilon(x+y)\nonumber\\
&&-{\textstyle\frac{1}{24}}(x^3+y^3)\,\varepsilon(x+y)-{\textstyle\frac{1}{24}}
(y^3-x^3)\,\varepsilon(y-x)\nonumber\\
&&-{\textstyle\frac{1}{4}}xy\{|x|[\theta(x-y)\,\theta(-x-y)+\theta(y-x)\,\theta
(x+y)]\nonumber\\
&&+|y|[\theta(y-x)\,\theta(-x-y)+\theta(x-y)\,\theta(x+y)]\}.
\label{e31}
\end{eqnarray}
We have plotted the function $\cC^{(2)}(x,y)$ on the symmetric domain $-\frac{
\pi}{2}<(x,y)<\frac{\pi}{2}$ in Fig.~\ref{f2}.

In summary, our final result for the $\cC$ operator to order $\epsilon^2$ on
the symmetric domain $-\frac{\pi}{2}<(x,y)<\frac{\pi}{2}$ is given by
\begin{equation}
\cC(x,y)=\delta(x+y)+\epsilon\cC^{(1)}(x,y)+\epsilon^2\cC^{(2)}(x,y)+
\mathcal{O}(\epsilon^3),
\label{e32}
\end{equation}
where $\cC^{(1)}(x,y)$ is given in (\ref{e27}) and $\cC^{(1)}(x,y)$ is given
in (\ref{e31}). We have verified by explicit calculation that to order
$\epsilon^2$ this $\cC$ operator obeys the algebraic equations (\ref{e8}).
For example, in coordinate space the third of these equations, $\cC^2=1$, reads
\begin{equation}
\int_{-\pi/2}^{\pi/2}dy\,\cC(x,y)\,\cC(y,z)=\delta(x-z)+\mathcal{O}(\epsilon^3).
\label{e33}
\end{equation}

\subsection{Calculation of the $Q$ operator}
The last step in the calculation is to determine the operator $Q$ from the
result in (\ref{e32}) by using (\ref{e9}). This is a long and difficult
calculation: We first multiply $\cC(x,z)$ on the right by $\delta(z+y)$, the
parity operator $\cP$ in coordinate space, and then integrate with respect to
$z$. This gives the coordinate space representation of $\cC\cP=e^Q$. Next, we
take the logarithm of the resulting expression and expand it as a series in
powers of $\epsilon$ to obtain $Q$. We find that the coefficient of $\epsilon^2$
in this expansion is zero, and thus we obtain the simple result in (\ref{e11}),
which is the principal result in this paper.

\section{Conclusions}
\label{s3}

In this paper we have used perturbative methods to calculate the $\cC$ operator
to second order in powers of $\epsilon$ for the complex $\cP\cT$-symmetric
square-well potential (\ref{e2}). Expressing $\cC$ in the form $e^Q\cP$, we have
found that the operator $Q$ for this model has an expansion in odd powers of
$\epsilon$, just as in the case of the cubic $\cP\cT$-symmetric oscillator whose
Hamiltonian is given in (\ref{e7}). Our result (\ref{e11}) for $Q$ is an
elementary function. We have verified our calculation of the $\cC$ operator
by showing that it satisfies the algebraic conditions (\ref{e8}).

The most noteworthy property of the $\cC$ operator is that the associated
operator $Q$ is a nonpolynomial function, and this kind of structure had not
been seen in previous studies of $\cC$. At the beginning of this calculation we
expected that for such a simple $\cP\cT$-symmetric Hamiltonian it would be
possible to calculate the $\cC$ operator exactly and in closed form. We find it
surprising that even for this elementary model the $\cC$ operator is so
nontrivial.

\vspace{0.5cm}
\begin{footnotesize}
\noindent
CMB is supported by the US Department of Energy.
\end{footnotesize}

\end{document}